\documentclass[12pt]{article}
\usepackage{graphicx}
\begin{document}
\centerline{\Large \bf The Sznajd Consensus Model with Continuous Opinions}

\bigskip
\bigskip

\centerline{Santo Fortunato}

\vskip0.3cm

\centerline{Fakult\"at f\"ur Physik, Universit\"at Bielefeld, 
D-33501 Bielefeld, Germany}

\noindent

\centerline{{\tt e-mail: fortunat@physik.uni-bielefeld.de}}

\bigskip

\begin{abstract}

In the consensus model of Sznajd, opinions are integers and a randomly 
chosen pair of neighbouring agents with the same opinion forces 
all their neighbours to share that 
opinion. We propose a simple extension of the model to continuous opinions,
based on the criterion of bounded confidence which is at the basis of 
other popular consensus models. Here the opinion $s$ is a real number between $0$
and $1$, and a parameter $\epsilon$ is introduced such that two  
agents are compatible if their opinions differ from each other by less than $\epsilon$.
If two neighbouring agents are compatible, they take the mean $s_m$ of their opinions
and try to impose this value to their neighbours. 
We find that if all neighbours take the average opinion $s_m$
the system reaches complete consensus for any
value of the confidence bound $\epsilon$. We propose as well a weaker 
prescription for the dynamics and discuss the corresponding results.

\end{abstract}

\bigskip

Keywords: Sociophysics, Monte Carlo simulations.

\bigskip

\section{Introduction}

\bigskip

The consensus model of Sznajd \cite{Sznajd} has rapidly acquired importance in the 
new field of computational sociophysics \cite{weidlich,staufrev}, where 
one tries to model society as a system of agents which interact
with each other, with the ultimate aim
to explain the occurrence at a global level of complex phenomena like the
formation of hierarchies
\cite{bonabeau} and consensus \cite{Sznajd, Axel, Deff, HK, Galam}.
 
In the original formulation of Sznajd, the agents sit on the sites of
a one-dimensional chain, and the opinion variable $s$ can take only the values
$\pm{1}$ ('up' or 'down'). In this respect the Sznajd model can be considered the "Ising model of
opinion dynamics". 
Initially each agent takes opinion $+1$ with probability $p$ and $-1$
with probability $1-p$. The dynamics is based on the principle that if two 
friends share the same opinion, they may succeed in convincing their
acquaintances of their opinion ("united we stand, divided we fall"). 
In the Sznajd algorithm, one randomly chooses a pair 
of neighbouring agents $i$ and $i+1$ and check whether their opinions $s_i$ and
$s_{i+1}$ are the
same ($++$ or $--$). If this is the case, 
their neighbours $i-1$ and $i+2$ take the opinion of $i$ and $i+1$
(so one finally has $s_{i-1}=s_{i}=s_{i+1}=s_{i+2}$).
It may of course happen that $s_i{\neq}s_{i+1}$ ($+-$ or $-+$). 
In this case, each agent of the pair "imposes" its opinion to the 
neighbour of the other agent of the pair, so $s_{i-1}=s_{i+1}$ and
$s_{i+2}=s_i$.
This second rule has usually been neglected in the successive studies on the
Sznajd model. In these works one used the so-called "basic" Sznajd dynamics,
where the opinions of the neighbours of the chosen pair of agents change
only if the two agents agree, otherwise nothing happens.
In its basic version, the Sznajd dynamics leads to a configuration where 
all agents share the same opinion (consensus), for any value of the initial
concentration $p$ of (up) opinions. If $p<1/2$ ($>1/2$) all agents will have opinion
$-1$ ($+1$) in
the final configuration. 

In this paper we will mainly deal with the basic Sznajd dynamics, but we will
as well present interesting results corresponding to the original Sznajd prescription.
Meanwhile a lot is known on this model. A great deal of refinements have been 
introduced, which can be grouped in two categories: variations of the  
social topology and modifications of the "convincing" rule.
Several lattice topologies have been adopted, simple square \cite{stauf1}, 
cubic \cite{kertesz}, triangular \cite{chang}, dilute \cite{moreira}, etc.
Moreover, network topologies have also been investigated, like pseudo-fractal 
\cite{gonzales} and especially 
scale free networks \cite{kertesz,sousa1}, which are currently very popular
\cite{netw}.
As far as the dynamics is concerned, one has explored 
what happens when $i$ neighbouring sites ($i{\geq}1$), not necessarily two, 
convince their neighbours, for the cases $i=1$ (single site) \cite{gonzales,ochr}
and $i=3$ \cite{gonzales,sousa1}.
Furthermore, one has also studied the case where the possible number of 
opinion states is larger than two \cite{kertesz,gonzales,bern,bonne,stauf2}.
The interest in the Sznajd model is not simply academic, as with this model
one was able to reproduce the distribution of the number of candidates    
according to the number of votes they received in Brazilian and Indian
elections \cite{kertesz,gonzales}.

Here we do not want to concentrate on specific applications or refinements of
the model, but rather reformulate it for the case in which opinions are real
numbers.
There are two reasons why this formulation could be important:

\begin{itemize}
\item{it deals with the case in which each individual has, at least initially,
    its own attitude/opinion, so one does not have to introduce the total number
  of possible opinions as a parameter;}
\item{it allows a direct comparison of the Sznajd dynamics and its predictions with 
the other two consensus models with real opinions, that of Deffuant et al. 
\cite{Deff} and that of Krause-Hegselmann (KH) \cite{HK}.} 
\end{itemize}

We start from a society where the relationships between the people 
are represented by the edges of a graph, not necessarily a regular lattice. 
The first step of the algorithm consists in assigning to each
agent a real number between $0$ and $1$ with uniform
probability. After that, as in the prescription of Sznajd, we choose a pair 
of neighbouring agents ($i$, $j$) and compare their opinions
$s_i$ and $s_j$. This is the point where we
need to introduce a new prescription. The opinions, being real, can never be
equal, as required by the Sznajd rule, but we have to soften this condition. 
As a matter of fact, instead of equality, we can demand "closeness", i.e. that
the two opinions must differ from each other by less than some real number $\epsilon$.  
This immediately recalls the principle of Bounded Confidence which characterizes 
both the model of Deffuant et al. and that of KH. There the parameter $\epsilon$
is called confidence bound and, if $|s_i-s_j|<\epsilon$, the two agents 
are compatible, in the sense that their positions are close enough to 
allow a discussion (interaction) between them; the discussion 
leads to a modification of their positions. In our case, 
we shall keep the denomination of confidence bound for $\epsilon$, but
the concept acquires a
slightly stronger meaning: we say that if $|s_i-s_j|<\epsilon$ the two agents 
are compatible enough to share the same opinion $s_m$ after their interaction, 
where $s_m=(s_i+s_j)/2$. This is actually what happens in the Deffuant model 
when the so-called convergence parameter $\mu=1/2$ \cite{Deff}.
If instead $i$ and $j$ are not compatible, both $i$ and $j$ maintain their
opinions $s_i$ and $s_j$.

Now we must define what happens to the opinions of the neighbours of the pair
$(i,j)$. If $i$ and $j$ are not compatible, we do nothing, as in the basic
version of Sznajd we mentioned above.
If $i$ and $j$ are compatible, 
we devise two possible prescriptions, that we call "Strong Continuous (SC) Sznajd"
and "Weak Continuous (WC) Sznajd" such that:

\begin{itemize}
\item{in SC Sznajd, all neighbours take the opinion $s_m$ of the pair,
    independently of their own opinions;}
\item{in WC Sznajd, only the agents which are compatible with
their neighbour in the pair $(i,j)$ take the opinion $s_m$, where the
compatibility refers to the opinion of the neighbour center site $i$ or $j$ 
before it gets updated to $s_m$.}
\end{itemize}

We shall see that these two prescriptions lead to very different results.
We update the opinions of the agents in the following way: we make an ordered sweep
over all agents, and, for each agent $i$, we  
select at random one of its neighbours and apply our version of 
the Sznajd prescription.
We repeat the procedure over and over until we find that, after a sweep, 
the opinion of each agent did not change appreciably, where "appreciably" for us
means by more than $10^{-9}$. 
We remark that in all studies on the Sznajd model one usually performed  
random and not sequential updates: for this reason we made some tests with random
updating, and the results are the same for SC Sznajd and essentially the same
for WC Sznajd.   
In all simulations we adopted two kinds of graphs
to describe society, a square lattice with periodic boundary conditions
and a Barab{\'a}si-Albert (BA) network \cite{BA}. 
A BA network with $N$ vertices can be constructed with a simple dynamical procedure.
First one has to 
specify the outdegree $m$ of the vertices, i.e. the number of edges which originate
from a vertex. One starts from $m$ vertices which are all
connected to each other and adds further $N-m$ vertices one at a time.
When a new vertex is added, it selects $m$ of the pre-existing vertices as neighbours,
so that the probability to get linked to a vertex is proportional
to the number of its neighbours.

\begin{figure}[hbt]
\begin{center}
\includegraphics[angle=-90,scale=0.55]{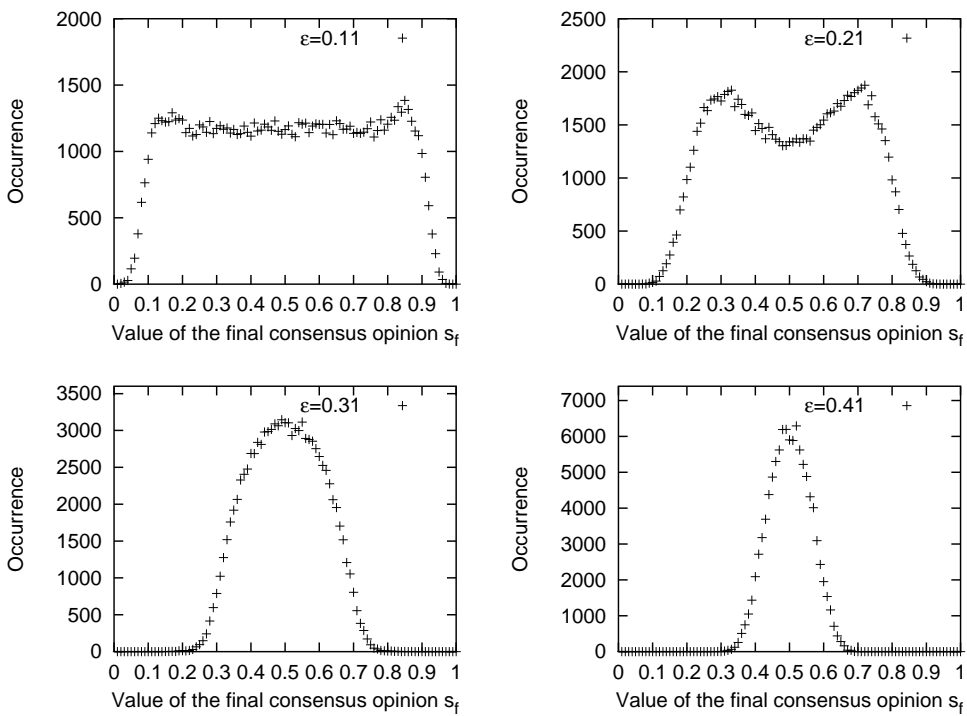}
\end{center}
\caption{\label{fig1} Probability distribution of the final surviving opinion for 
Strong Continuous Sznajd. The social topology is a square lattice with $2500$ sites.}
\end{figure}

Since one needs to fix the value of the confidence bound $\epsilon$ before 
starting the simulation, the results will in general depend on $\epsilon$ and we
shall investigate this dependence. 
Let us start to present the results relative to SC Sznajd. In all simulations we
have carried on, both on the lattice and on BA networks, we found that the
system converges to a configuration where all agents have one and the same
opinion (complete consensus), for any value of $\epsilon$. This result,
which matches that of the original discrete version, 
shows that the Sznajd dynamics is most effective to achieve 
a full synchronization of the agents. We remark that the result holds 
independently of the initial distribution of opinions, which needs not be uniform. 
We also found that the value of the 
final opinion $s_f$ of the agents is not $1/2$, as in the models of Deffuant and KH,
but it can take any value in a range centered at $1/2$. The width of the range
and the probability distribution of $s_f$ depend on $\epsilon$. In Fig. \ref{fig1}
we show the probability distribution of $s_f$ for 
a square lattice and four values of $\epsilon$, 
obtained from $100000$ runs.
As one can see, the histograms are all symmetric with respect to the center
opinion $1/2$, as expected, but their shape varies with $\epsilon$.
We distinguish three characteristic profiles, flat, double peaked 
and single peaked for low, intermediate and high values of $\epsilon$,
respectively. In the case of a single peak, we have noticed that the width 
shrinks approximately as $1/\sqrt{N}$, when $N$ increases; so the peak is 
probably doomed to become a $\delta$-function centered at $1/2$ 
when $N\rightarrow\infty$. On the other hand, at low $\epsilon$, we noticed that 
the histogram does not change appreciably when $N$ increases. 
This means that there must be some
$\epsilon_c$ such that if $\epsilon<\epsilon_c$ the final opinion $s_f$ falls
in a finite range of opinions, if instead $\epsilon>\epsilon_c$ $s_f=1/2$.

\begin{figure}[hbt]
\begin{center}
\includegraphics[angle=-90,scale=0.50]{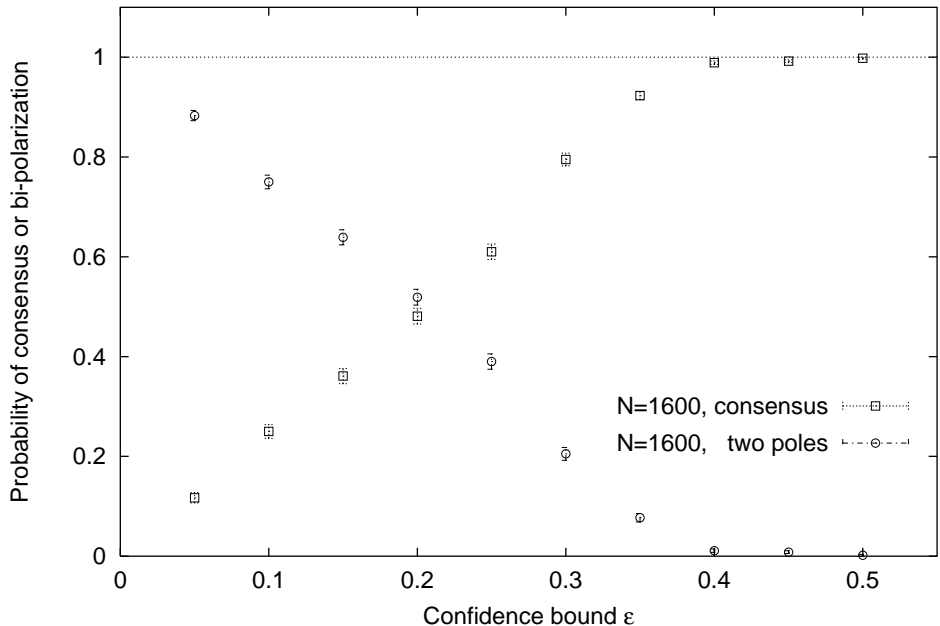}
\end{center}
\caption{\label{fig2}Fraction of samples with complete consensus and
  bi-polarization for Strong Continuous Sznajd with both "ferromagnetic" and 
"anti-ferromagnetic"
coupling. The agents sit on the sites of a square lattice of side $L=40$.}
\end{figure}

\begin{figure}[hbt]
\begin{center}
\includegraphics[angle=-90,scale=0.50]{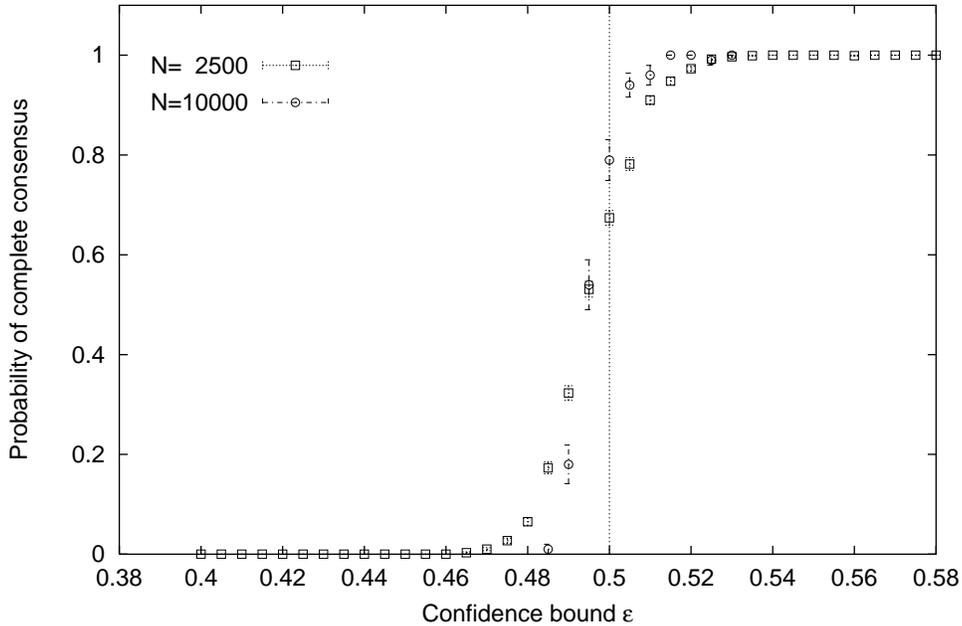}
\end{center}
\caption{\label{fig3}Fraction of samples with complete consensus as a function 
of $\epsilon$, for Weak Continuous Sznajd on two square lattices with
$2500$ and $10000$ agents.}
\end{figure}

\begin{figure}[hbt]
\begin{center}
\includegraphics[angle=-90,scale=0.50]{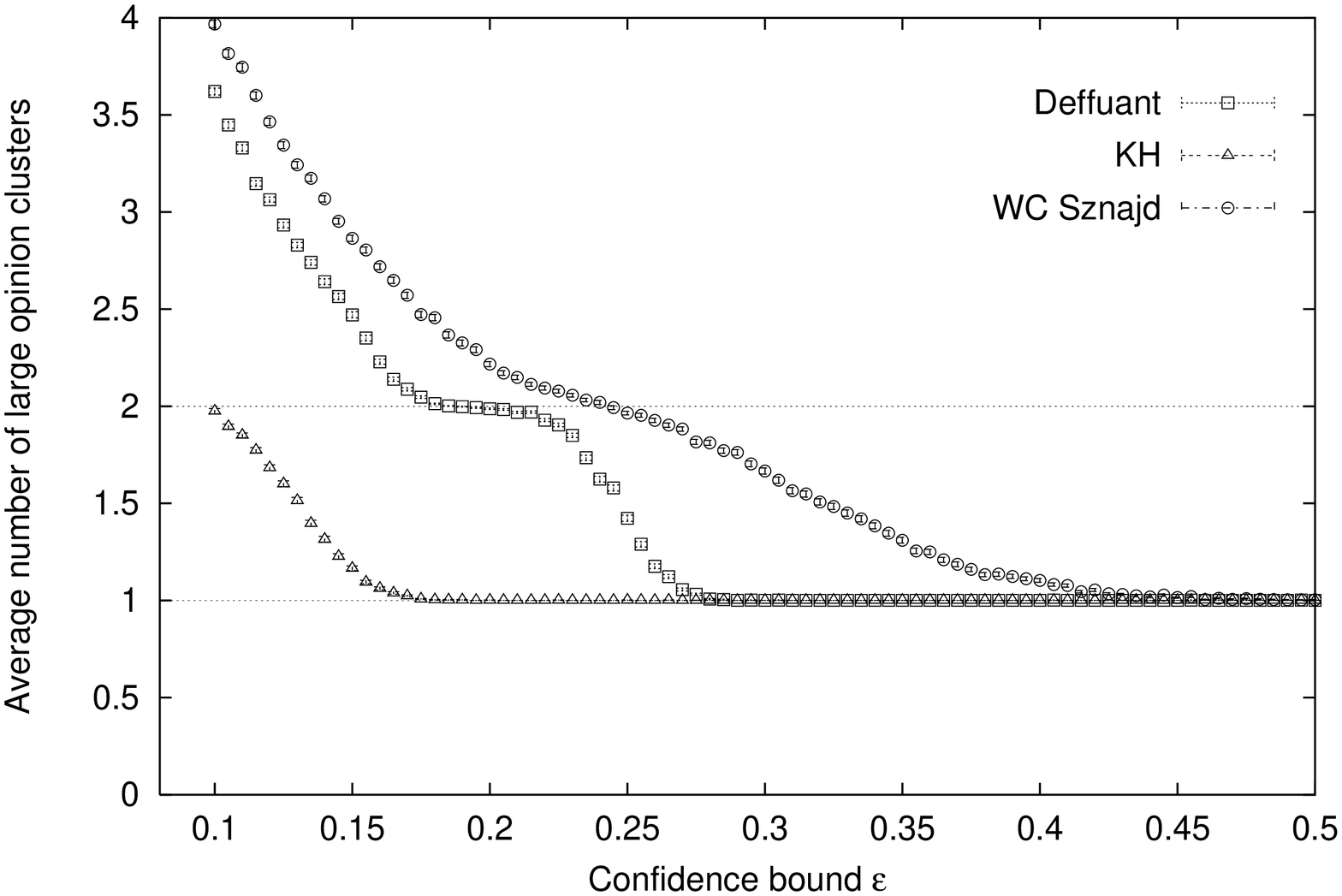}
\end{center}
\caption{\label{fig4}Average number of large opinion clusters as a function 
of $\epsilon$ for Weak Continuous Sznajd, Deffuant and Krause-Hegselmann. 
The social topology is a 
scale free network {\'a} la Barab{\'a}si-Albert with 1000 vertices.}
\end{figure}

Since the structure of the final opinion configuration is always
the same, i.e. consensus, we checked what happens if we add to the convincing rule of
the basic Sznajd dynamics the "anti-ferromagnetic" prescription originally
introduced in the seminal paper \cite{Sznajd}, for the case in which the
opinions of the agents of the randomly selected pair $(i,j)$ are not compatible. 
In this case, the extension to our case is trivial: the neighbours of $i$
take the opinion of $j$ and viceversa. The effect of this more complex dynamics
is that the system can converge to one of only two possible situations: either
there is complete consensus, like before, or there is a perfect splitting 
of the community in two factions, with exactly half of the agents sharing either opinion.
In this case, too, one confirms the result obtained with the discrete Sznajd
model, where one would have either a perfect ferromagnet (all agents
with opinions $+1$ or $-1$), or a perfect antiferromagnet (with the 
opinions $+1$ and $-1$ which regularly alternate in the chain/lattice).
Indeed, when the population splits in two factions, the two opinions
regularly alternate on the lattice, as this is the only possible
stable situation different from consensus.
In Fig. \ref{fig2} we plot the probability of having either of the final states,
i.e. the fraction of samples in which we obtained consensus or bi-polarization,
for different values of $\epsilon$. Society is a square lattice 
and the total number of samples is $1000$. 
We notice that bi-polarization is very likely to occur at low values of
$\epsilon$, whereas one always obtains consensus for $\epsilon$ larger than 
about $0.4$ (although the real threshold is probably $1/2$, as in the model 
of Deffuant \cite{santo}). 

Now we turn to Weak Continuous Sznajd. We believe that this is a more 
realistic implementation of the Sznajd dynamics, as only people whose
positions are somewhat close to each other can be influenced.
The fact that we apply bounded confidence
to the neighbours as well dramatically changes the scenario. Now one can have 
a variable number of opinion clusters in the final configuration, 
depending on the value of the confidence bound, as in the
models of Deffuant and KH. First, we tried to determine the threshold for
complete consensus. For this purpose we calculated again the fraction of samples
with a single surviving opinion, out of $1000$ total configurations, for several values
of $\epsilon$. Fig. \ref{fig3} shows the results, where we took again 
a square lattice topology and two different sizes to investigate the  
limit when the number of agents goes to infinity. From the figure
it is clear that the threshold for complete consensus is $1/2$, as in the model
of Deffuant \cite{santo}. A similar analysis on Barab{\'a}si-Albert networks
confirms the result. 

Next, we compared the model with the other two continuous opinion models,
Deffuant and KH. One of the most important issues is the variation with
$\epsilon$ of the number of clusters in the final configuration. We decided to
focus on large clusters: we say that a cluster is large if it includes
more than $\sqrt{N}$ agents, where $N$ is as usual the size of the total population.
As a matter of fact, especially when $N$ is not too large, as in the cases we
have examined, it quite often happens that in the final configuration
several clusters with very few agents co-exist with larger ones. Most small clusters
are artefacts due to the finite size of the system, and would disappear if $N$
becomes large. That is why we focus on large clusters, which represent most of
the real
parties/factions created by the dynamics in the limit $N\rightarrow\infty$.
Fig. \ref{fig4} shows the pattern of the large cluster multiplicity with
$\epsilon$, for WC Sznajd, Deffuant and KH, respectively. The system is a scale
free network {\'a} la Barab{\'a}si-Albert, with $1000$ vertices. Further
simulations at larger $N$ indicate that the pattern shown 
in the figure is nearly asymptotic, i.e. does not change appreciably when $N$ increases.
We see that there is a sort of monotonic relationship between the three models:
for a given $\epsilon$ there are more large clusters in the final configuration 
for WC Sznajd than for Deffuant, and more for Deffuant than for KH.
In particular one has to go to much higher values of $\epsilon$ for WC Sznajd
in order to obtain a single large cluster in the final configuration, a
situation
which is instead much easier to reach for the other two models. 

In conclusion, we have presented a generalization of the Sznajd dynamics to 
real-valued opinions, based on bounded confidence. 
We proposed two prescriptions for updating the opinions, 
which differ from each other by the influence of the randomly selected pair of
(compatible) agents on their neighbours. According to the first rule, 
all neighbours accept the average opinion of the pair. In this
case, the fate of the system is simple: all agents will end up with the same
opinion at some stage. The second rule, instead, limits the influence of the 
pair only to those neighbours which are compatible with their
friend in the pair. In this case one can have any number of opinion clusters in
the final configuration, depending on $\epsilon$, and consensus is attained only
for $\epsilon>1/2$. The latter prescription turns out to be 
less effective to create
large opinion clusters than the dynamics of Deffuant and Krause-Hegselmann.

\bigskip

I thank D. Stauffer for a critical reading of the manuscript. 
I gratefully acknowledge the financial support of the DFG Forschergruppe
under grant FOR 339/2-1.


\begin{thebibliography} {99}

\bibitem{Sznajd} K. Sznajd-Weron and J. Sznajd, Int. J. Mod. Phys. C {\bf 11},
  1157 (2000).

\bibitem{weidlich} W. Weidlich, {\it Sociodynamics; A Systematic Approach to 
Mathematical Modelling in the Social Sciences}. Harwood Academic Publishers,
2000.

\bibitem{staufrev} D. Stauffer, {\it The Monte Carlo Method on the Physical Sciences},
  edited by J. E. Gubernatis, 
AIP Conf. Proc. {\bf 690}, 147 (2003), cond-mat/0307133.

\bibitem{bonabeau} E. Bonabeau, G. Theraulaz and J. L. Deneubourg, 
Physica A {\bf 217}, 373 (1995).

\bibitem{Axel} R. Axelrod, J. Conflict Resolut. {\bf 41}, 203 (1997).

\bibitem{Deff} G. Deffuant, D. Neau, F. Amblard and G. Weisbuch,
Adv. Complex Syst. {\bf 3}, 87 (2000); G. Weisbuch, G. Deffuant, F. Amblard, and
J.-P. Nadal, Complexity {\bf 7}, 2002; G. Deffuant, F. Amblard, G. Weisbuch and 
T. Faure, Journal of Artificial Societies and Social Simulations {\bf 5}, issue
4, paper 1 (jasss.soc.surrey.ac.uk) (2002).

\bibitem{HK} R. Hegselmann and U. Krause, Journal of Artificial Societies and 
Social Simulation {\bf 5}, issue 3, paper 2 (jasss.soc.surrey.ac.uk) (2002) and
Physics A, in press (2004); U. Krause, {\it Soziale Dynamiken 
mit vielen interakteuren. Eine Problemskizze}.
In U. Krause and
M. St{\"o}ckler (Eds.), {\it Modellierung und
Simulation von Dynamiken mit vielen interagierenden Akteuren}, 37-51,
Bremen University, Jan. 1997.

\bibitem{Galam} S. Galam, J. Stat. Phys. {\bf 61}, 943 (1990) and Physica A 
  {\bf 238}, 66 (1997).

\bibitem{stauf1} D. Stauffer, A. O. Sousa and S. Moss de Oliveira,
  Int. J. Mod. Phys. C {\bf 11}, 1239 (2000).

\bibitem{kertesz} A. T. Bernardes, D. Stauffer and J. Kert{\'e}sz, 
Eur. Phys. J. B {\bf 25}, 123 (2002), cond-mat/0111147.

\bibitem{chang} I. Chang, Int. J. Mod. Phys. C {\bf 12}, 1509 (2001).

\bibitem{moreira} A. A. Moreira, J. S. Jr. Andrade and D. Stauffer,
  Int. J. Mod. Phys. C {\bf 12}, 39 (2001). 

\bibitem{gonzales} M. C. Gonzalez, A. O. Sousa and H. J. Herrmann, 
Int. J. Mod. Phys. C {\bf 15}, 45 (2004), cond-mat/0307537.

\bibitem{sousa1} A. O. Sousa, cond-mat/0406390.

\bibitem{netw} R. Albert and A. L. Barab{\'a}si, Rev. Mod. Phys. {\bf 74}, 47 (2002);
M. E. J. Newman, SIAM Review 45, 167 (2003).

\bibitem{ochr} R. Ochrombel, Int. J. Mod. Phys. C {\bf 12}, 1091 (2001).

\bibitem{bern} A. T. Bernardes, U. M. S. Costa, A. D. Araujo and D. Stauffer, 
Int. J. Mod. Phys. C {\bf 12}, 159 (2001).

\bibitem{sta2} D. Stauffer, A. O. Sousa and C. Schulze, Journal of Artificial
Societies and Social Simulation {\bf 7}, issue 3, paper 7 (2004). 

\bibitem{bonne} J. Bonnekoh, Int. J. Mod. Phys. C {\bf 14}, 1231 (2003).

\bibitem{stauf2} D. Stauffer, Adv. Compl. Syst. {\bf 5}, 97 (2002) and 
Int. J. Mod. Phys. C {\bf 13}, 315 (2002).

\bibitem{BA} A. L. Barab{\'a}si and R. Albert, Science {\bf 286}, 509 (1999).

\bibitem{santo} S. Fortunato, cond-mat/0406054.

\end{thebibliography}
\end{document}